\documentclass[a4paper,nobibnotes,nofootinbib,notitlepage,superscriptaddress]{revtex4}

\usepackage{amssymb}
\usepackage{amsmath}
\usepackage{epsfig}

\newcommand{\be}{\begin{equation}}
\newcommand{\ee}{\end{equation}}
\newcommand{\bea}{\begin{eqnarray}}
\newcommand{\eea}{\end{eqnarray}}

\newcommand{\pom}{\mathbb{P}}
\newcommand{\reg}{\mathbb{R}}

\begin{document}

\title{Probing the Pomeron structure using dijets and 
$\gamma$+jet events at the LHC}

\author{C. Marquet}\email{cyrille.marquet@cern.ch}
\affiliation{Centre de physique th\'eorique, \'Ecole Polytechnique, CNRS, 91128 Palaiseau, France}

\author{C. Royon}\email{christophe.royon@cea.fr}
\affiliation{IRFU/Service de Physique des Particules, CEA/Saclay, 91191 Gif-sur-Yvette cedex, France}

\author{M. Saimpert}\email{matthias.saimpert@cea.fr}
\affiliation{IRFU/Service de Physique des Particules, CEA/Saclay, 91191 Gif-sur-Yvette cedex, France}

\author{D. Werder}\email{dominik.werder@physics.uu.se}
\affiliation{Department of Physics and Astronomy, Uppsala University, Box 516, SE-751 20 Uppsala, Sweden}

\begin{abstract}

We consider hard diffractive events in proton-proton collisions at the LHC, 
in which both protons escape the collision intact. In such double Pomeron 
exchange processes, we propose to measure dijets and photon-jet final states, and we show 
that it has the potential to pin down the Pomeron quark and gluon contents, a crucial 
ingredient in the standard QCD description of hard diffraction. By comparing 
with predictions of the soft color interaction approach, we also show that 
more generally, the measurement of the photon-jet to di-jet cross section 
ratio can put stringent test on the QCD dynamics at play in diffractive 
processes in hadronic collisions.

\end{abstract}

\maketitle

\section{Introduction}

Understanding diffractive interactions in QCD has been a challenge for many years. In particular, for hard diffractive events in hadronic collisions, it seems that a description based solely on weak-coupling methods cannot be obtained. In the case of deep inelastic scattering (DIS), due to several years of experimental efforts at HERA, as well as many theoretical achievements, the situation has reached a satisfactory level. The diffractive part of the deep inelastic cross-section is now well understood in perturbative 
QCD \cite{Chekanov:2008cw,Chekanov:2008fh,Aaron:2010aa,Aaron:2012ad}. This is achieved by means of different approximation schemes each valid in their own kinematic limit. For instance, at large photon virtuality $Q^2$, one can use the collinear factorization of diffractive parton densities \cite{Collins:1997sr}, while at small values Bjorken x, the non-linear evolution of dipole scattering amplitudes is also successful \cite{GolecBiernat:1999qd,Kovchegov:1999ji,Marquet:2007nf}. By contrast, the description of hard diffraction in hadron-hadron collisions still poses great theoretical challenges.

Tevatron data provided evidence that, even at very large scales, collinear factorization does not apply for diffraction in hadron-hadron collisions \cite{Affolder:2000vb}. There exist however empirical indications that the factorization breaking can be compensated by an overall factor, called the gap survival probability, independent of the details of the hard process. Nevertheless, many questions remain unanswered. Is this factor only a function of the collision energy, as often assumed? Does one need a different factor in single diffraction, double diffraction, and double-Pomeron-exchange (DPE) processes? One should be able to answer these questions at the LHC, where diffraction is a large part of the QCD program \cite{afp}. In addition, should this picture of diffraction be validated, it will be possible to further constraint the details of the theoretical description.

In this paper we focus on one aspect of this picture of diffractive events: the so-called Regge factorization of the diffractive parton densities into a Pomeron flux and Pomeron parton distributions. Such factorization is successful in DIS, but whether it also works in hadron-hadron collisions remains to be proven. In particular, the resulting Pomeron content in terms of quarks and gluons should be consistent with what has been extracted from HERA data. In order to constrain the Pomeron gluon and quark densities at the LHC, we shall consider the di-jet and photon-jet processes respectively, in DPE events, meaning that both protons escapes the collision intact. We show that the measurement of the photon-jet to di-jet cross section ratio has the potential to pin down the Pomeron quark content, or at the very least put stringent test on the Pomeron-like picture of diffraction.

On the contrary, a different mechanism could be responsible for diffraction in hadron-hadron collisions.
For this reason, we also compare the previous expectations for the photon-jet to di-jet cross section ratio in DPE processes with predictions from a different picture of diffraction,
which does not invoke Pomerons nor diffractive parton distributions:
the soft color interaction (SCI) model \cite{Edin:1995gi,Rathsman:1998tp}.
This approach models diffractive events exactly the same as other inelastic events,
but allows for additional soft exchanges which do not significantly change the kinematics
but change the color topology of the event.
This changed topology can then result in rapidity gaps and leading final state protons
after hadronization.
In this alternative description of diffractive interactions,
the QCD dynamics at play is very different compared to the more standard Pomeron picture.

The plan of the paper is as follows. In Section 2, we present more details about the 
theoretical description of hard diffractive events in hadron-hadron collisions, and on their implementation into the forward physics Monte Carlo (FPMC) program, which is used in our analysis. In Section 3, we discuss the sensitivity of the DPE di-jet process to the gluon content of the Pomeron, this will be the first DPE measurement performed at the LHC. In Section 4, we focus on the quark content of the Pomeron, and show the improvements that can 
be obtained by measuring DPE photon-jet processes. Finally, in Section 5, we compare our results with those obtained in the SCI model, for the photon-jet to di-jet cross section ratio. Section 6 is devoted to conclusions and outlook.

\section{Hard diffractive processes and their implementation in FPMC}

\begin{figure}[t]
\begin{center}
\epsfig{file=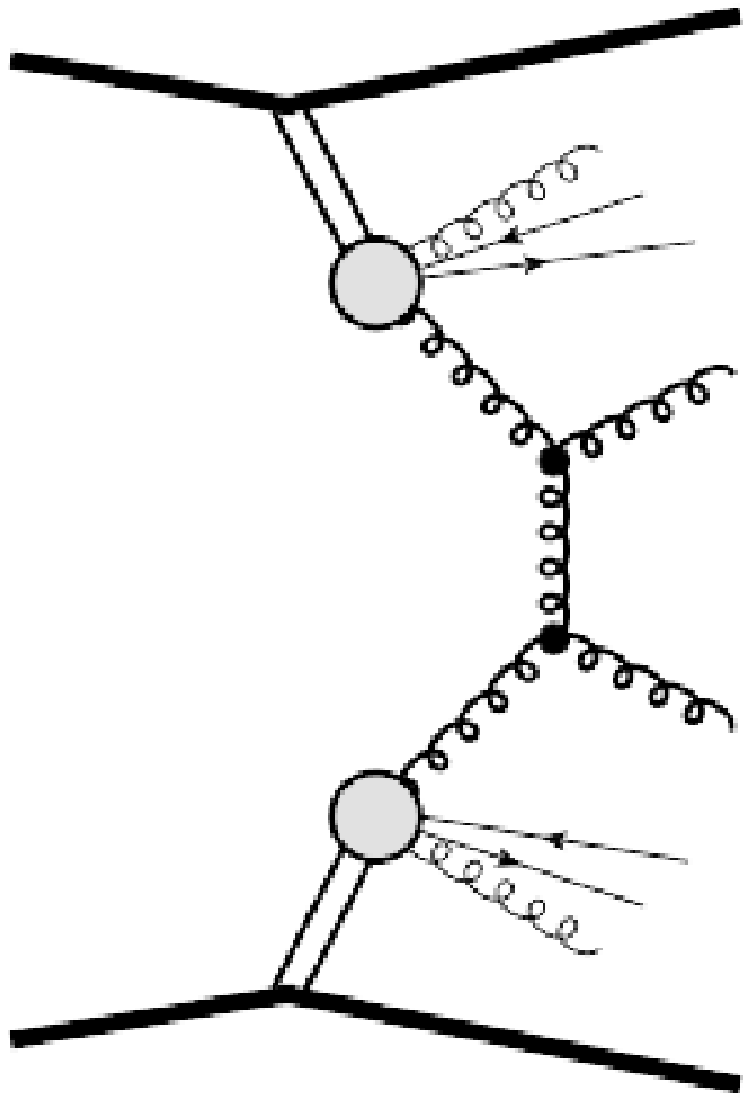,width=4.5cm}
\epsfig{file=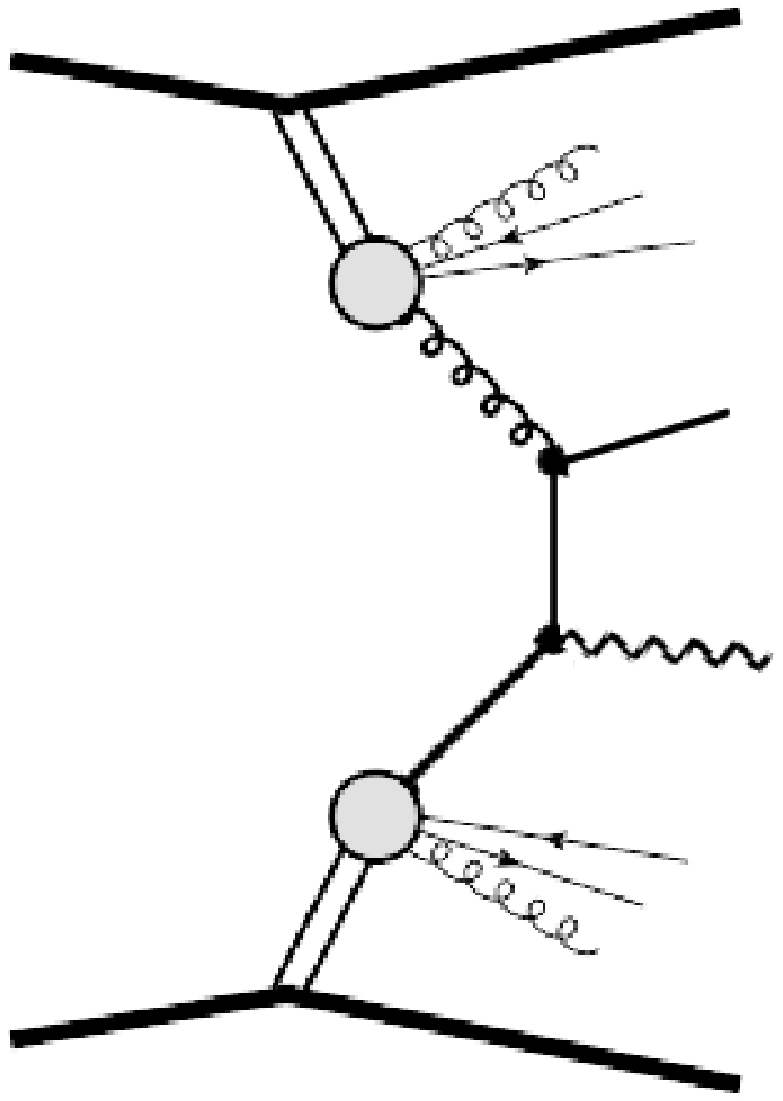,width=4.5cm}
\caption{Leading-order diagrams for double-Pomeron exchange di-jet (left) and $\gamma +$ jet (right) production in proton-proton collisions. The di-jet process is sensitive to the Pomeron gluon density and the $\gamma +$ jet process to the Pomeron quark densities.}
\label{fig0}
\end{center}
\end{figure}

We shall focus on double Pomeron exchange processes in proton-proton collisions. The
formulation of single- and double-diffractive processes is very similar, those cross
section can be obtained with simple modifications to what is presented in this section.
In addition, we write our formulae for the di-jet final state $J+J+X$, but they hold for
the photon-jet final state $\gamma+J +X$ as well. The leading-order diagrams for those
processes are pictured in Fig.~\ref{fig0}, and the following
long-distance/short-distance factorization formula is used to compute the cross
sections:
\be
d\sigma^{pp\to pJJXp}={\cal S}_{DPE}\ \sum_{i,j} \int d\beta_1d\beta_2\
f^D_{i/p}(\xi_1,t_1,\beta_1,\mu^2)f^D_{j/p}(\xi_2,t_2,\beta_2,\mu^2)\
d\hat\sigma^{ij\to JJX}
\label{colfact}
\ee
where $d\hat\sigma$ is the short-distance partonic cross-section, which can be computed
order by order in perturbation theory (provided the transverse momentum of the jets is
sufficiently large), and each factor $f^D_{i/p}$ denotes the diffractive parton
distribution in a proton. These are non-perturbative objects, however their evolution
with the factorization scale $\mu$ is obtained pertubatively using the
Dokshitzer-Gribov-Lipatov-Altarelli-Parisi~\cite{dglap} evolution equations. The
variables $\xi$ and $t$ denote, for each intact proton, their fractional energy loss and
the momentum squared transferred into the collision, respectively.

Hard diffractive cross sections in hadronic collisions do not obey collinear
factorization. This is due to possible secondary soft interactions between the colliding
hadrons which can fill the rapidity gap(s). Formula \eqref{colfact} is reminiscent of
such a factorization, but it is corrected with the so-called gap survival probability
{\cal S} which is supposed to account for the effects of the soft interactions. Since
those happen on much longer time scales compared to the hard process, they are modeled
by an overall factor. This is part of the assumptions that need to be further tested at
the LHC. To evaluate hard cross sections in the single diffraction case, one of the
diffractive parton distributions in \eqref{colfact} is replaced by a regular parton
distribution, and the survival probability factor ${\cal S}_{SD}$ is different as well.

To produce single diffractive and double Pomeron exchange events, FPMC uses diffractive
parton distributions extracted from HERA data~\cite{Aktas:2006hy} on diffractive DIS (a
process for which collinear factorization does hold). These are decomposed further into
Pomeron and Reggeon fluxes $f_{\pom,\reg/p}$ and parton distributions $f_{i/\pom,\reg}$
:
\be
f^D_{i/p}(\xi,t,\beta,\mu^2)= f_{\pom/p}(\xi,t)
f_{i/\pom}(\beta,\mu^2)+f_{\reg/p}(\xi,t) f_{i/\reg}(\beta,\mu^2)\quad\mbox{with}
\quad f_{\pom,\reg/p}(\xi,t)=\frac{e^{B_{\pom,\reg}t}}{\xi^{2\alpha_{\pom,\reg}(t)-1}}\
\ee

The diffractive slopes $B_{\pom,\reg}$, the Regge trajectories
$\alpha_{\pom,\reg}(t)=\alpha_{\pom,\reg}(0)+t\alpha'_{\pom,\reg}$ and the parton
densities $f_{i/\pom,\reg}$ can be found in the litterature. The secondary
reggeon contribution (with respect to the Pomeron one) has not yet been
implemented in the FPMC generator but it is assumed to occur at high $\xi$ at
the edge of the foward proton detector acceptance. Obviously, it is important to
measure at the LHC this contribution since the extrapolation at higher energies
of the present
secondary contribution performed by the H1 collaboration at HERA using the pion
structure function~\cite{Aaron:2012ad} is definietly questionable. A measurement of
the $\xi$ distribution in diffractive events for dijets for instance will be of
great interest since it should combine the $1/\xi$ dependence from the Pomeron
part with a flatter distribution at high $\xi$ originating from reggeons.

Measurements at the LHC will allow to test the validity of this further factorization of
the diffractive parton distributions into a Pomeron flux and Pomeron parton
distributions, as well as the universality of those Pomeron fluxes and parton
distributions. Finally, for those processes in which the partonic structure of the
Pomeron is probed, the existing HERWIG matrix elements of inelastic production are used
in FPMC to calculate the partonic cross sections, at leading order.

\section{Sensitivity to the Pomeron structure in gluons}
In this section, we detail the potential measurements to be performed at the LHC
in order to constrain the gluon structure in the Pomeron. We will start by
describing the experimental framework (the proton detectors to be installed
especially in the CMS and ATLAS experiments) and the Monte Carlo tools. We will
finish the section by giving the potential observables which will allow to
constrain the gluon content inside the Pomeron.
\subsection{Experimental framework}

In the following, we assume the intact protons to be tagged in the forward proton
detectors to be installed by the CMS/Totem and the ATLAS
collaborations~\cite{afp}. Two different places at 210-220 and 420 m in the LHC can be used to
install such detectors. The idea is to measure scattered protons at very small
angles at the interaction point and to use the LHC magnets as a spectrometer to
detect and measure the intact protons. In the following, we choose as an example
the acceptances of the forward detectors to be installed in the ATLAS
collaboration at 210 and 420m
\begin{itemize}
\item $0.015 < \xi < 0.15$ for 210 m detectors only
\item $0.0015 < \xi < 0.15$ for a combination of 210 and 420 m detectors.\
\end{itemize}
knowing the fact that the acceptance for such detectors in CMS-Totem is similar.

In order to produce Double Pomeron Exchange events, where both protons are intact
in the final state, and produce all studies 
mentioned in this paper we use the Forward Physics Monte Carlo (FPMC) 
generator~\cite{fpmc}.
FPMC aims to accommodate all relevant models for  
forward physics which could be studied at the LHC and contains in particular
the two-photon and double pomeron exchange processes. The generation of the 
forward processes is embedded 
inside HERWIG~\cite{herwig}. The great advantage of the program is that all processes 
with leading protons can be studied in the same framework, using the same 
hadronization model.

\subsection{Measurement of the dijet cross section and constraints on the gluon
density in the Pomeron}
\begin{figure}[t]
\begin{center}
\epsfig{file=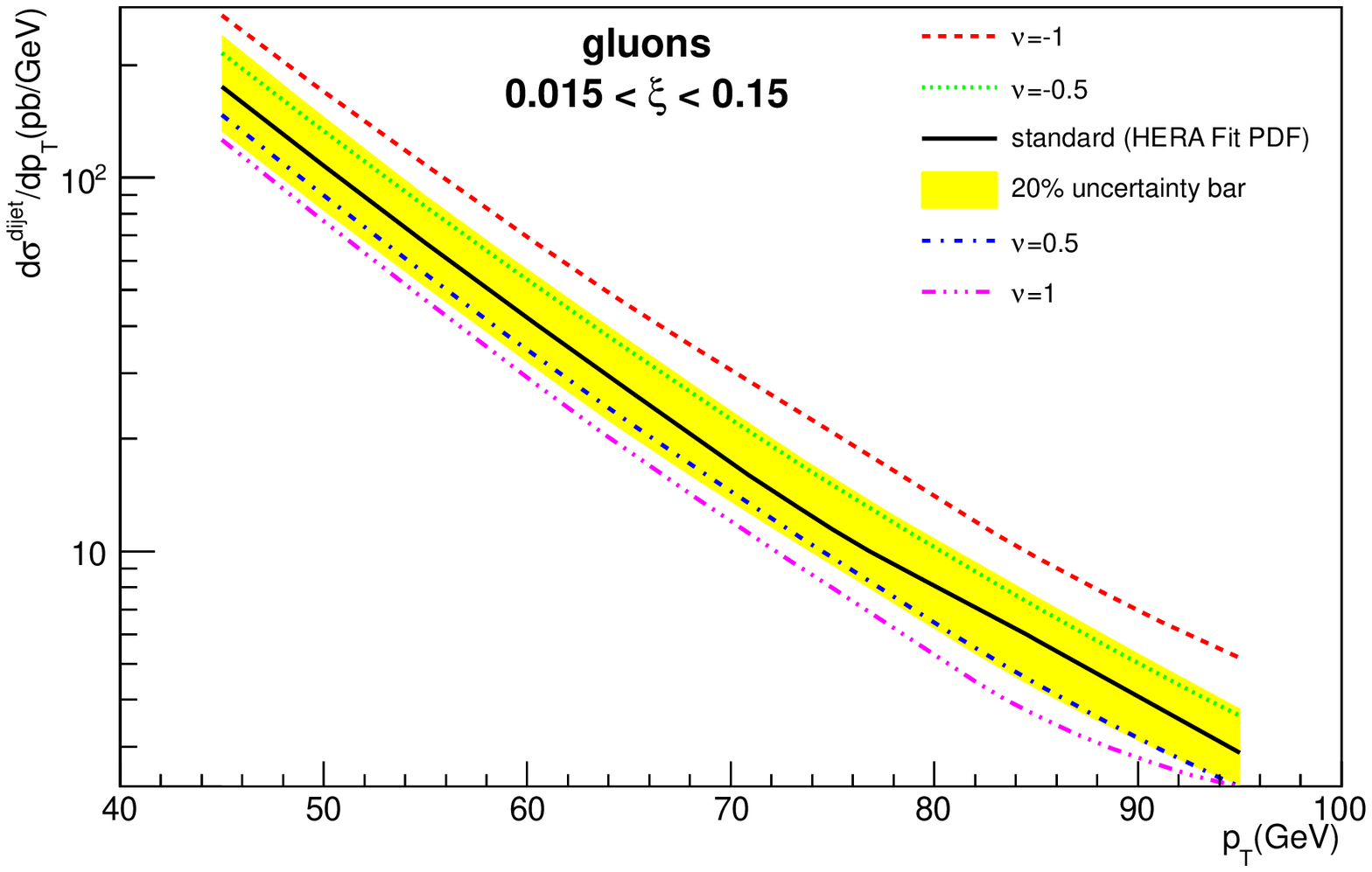,width=8.5cm}
\epsfig{file=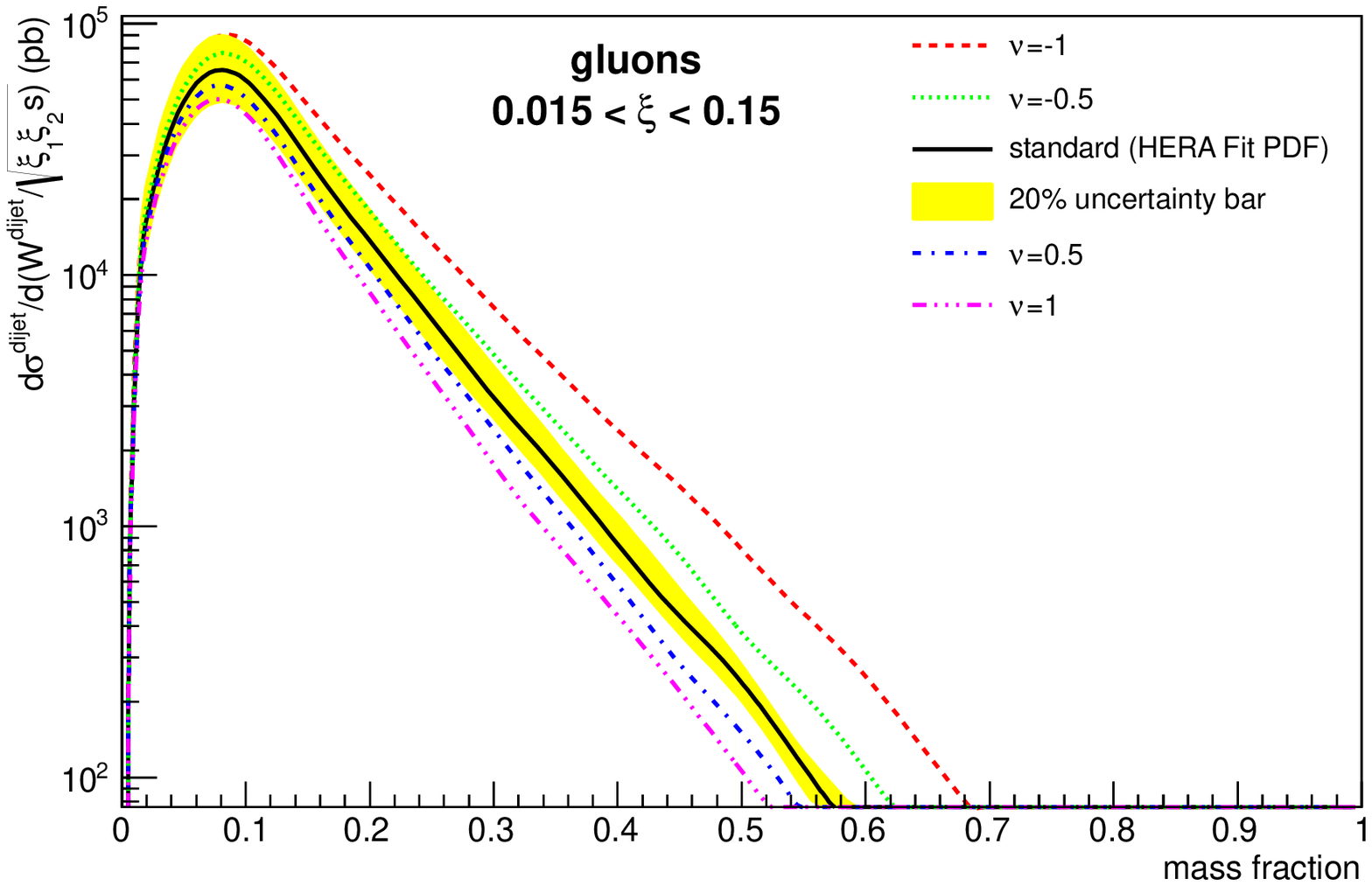,width=8.5cm}
\caption{Left: DPE di-jet cross section as a function of jet $p_T$ at the LHC. Right: DPE di-jet mass fraction distribution. The different curves correspond to different modifications of the Pomeron gluon density extracted from HERA data (see text).}
\label{fig6b}
\end{center}
\end{figure}

The dijet production in DPE events at the LHC is sensitive to the gluon density
in the Pomeron. The aim of this section is to study if we can constrain further
the gluon density in the Pomeron compared to the determination at HERA and to
check if the Pomeron model is universal between HERA and LHC. In order to
quantify how well we are sensitive to the Pomeron structure in terms of gluon
density at the LHC, we display in Fig.~\ref{fig6b}, left, the dijet cross
section as a function of the jet pT. The central black line displays the cross
section value for the gluon density in the Pomeron measured at HERA including an
additional survival probability of 0.03. The yellow band shows the effect of the
20\% uncertainty on the gluon density taking into account the normalisation
uncertainties. The dashed curves display how the dijet
cross section at the LHC is sensitive to the gluon density distribution
especially at high $\beta$. For this sake, we multiply the gluon density in the
Pomeron from HERA by $(1-\beta)^{\nu}$ where $\nu$ varies between -1 and 1. When
$\nu$ is equal to -1 (resp. 1), the gluon density is enhanced (resp, decreased)
at high $\beta$. From Fig.~\ref{fig6b}, we notice that the dijet cross section
is indeed sensitive to the gluon density in the Pomeron 
and we can definitely check if the Pomeron
model from HERA and its structure in terms of gluons is compatible between HERA
and the LHC. This will be an important test of the Pomeron universality. This
measurement can be performed for a luminosity as low as 10 pb$^{-1}$ since the
cross section is very large (typically, one day at low luminosity without pile
up at the LHC). It is worth noticing that this measurement will be
limited by systematic uncertainties (not the statistical ones). Typically, if
the jet energy scale is known with a precision of 1\%, we expect the systematics
on the jet cross section mainly due to jet energy scale and jet $p_T$ resolution
to be of the order of 15\%.

However, from this measurement alone, it
will be difficult to know if the potential difference between the expectations
from HERA and the measurement at the LHC are mainly due to the gluon density or
the survival probability since the ratio between the curves for the different
gluons (varying the $\nu$ parameters) are almost contant. It will be difficult
to know if these effects are rather due to the value of the survival
probability.

An additional observable more sensitive to the gluon density in the Pomeron is
displayed in Fig.~\ref{fig6b}, right. This is the so called dijet mass fraction,
the ratio of the dijet mass to the total diffractive mass computed as
$\sqrt{\xi_1 \xi_2 S}$ where $\xi_{1,2}$ are the proton fractional momentum
carried by each Pomeron and $S$ the center-of-mass energy of 14 TeV. We note
that the curves corresponding to the different values of $\nu$ are much more 
spaced at high values of the dijet mass fraction, meaning that this observable
is indeed sensitive to the gluon density at high $\beta$. This is due to the
fact that the dijet mass fraction is equal to $\sqrt{\beta_1 \beta_2}$
where $\beta_{1,2}$ are the Pomeron momentum fraction carried by the parton
inside the Pomeron which interacts. The measurement of the dijet cross section
as a function of the dijet mass fraction is thus sensitive to the product of the
gluon distribution taken at $\beta_1$ and $\beta_2$. It is worth mentioning tha
exclusive dijet events will contribute to this distribution at higher values of
the dijet mass fraction above 0.6-0.7~\cite{olda}.

Once the gluon distribution in the Pomeron will be constrained at the LHC using
dijet events and the measurement of the dijet mass fraction, it is possible to
constrain further the quark content inside the Pomeron.

\section{Sensitivity to the Pomeron structure in quarks using $\gamma +jet$ events}

\begin{figure}[t]
\begin{center}
\epsfig{file=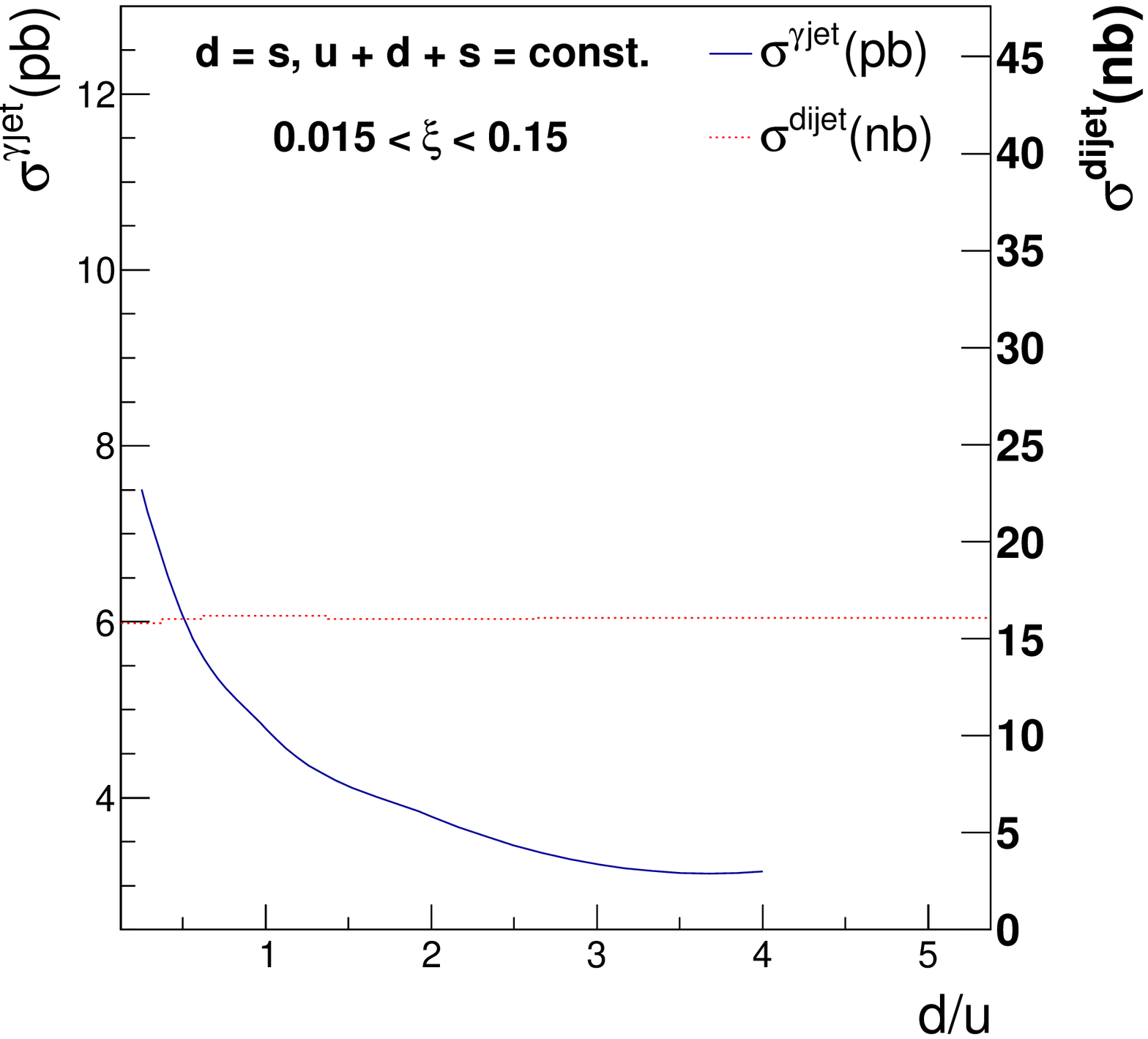,width=7cm}
\epsfig{file=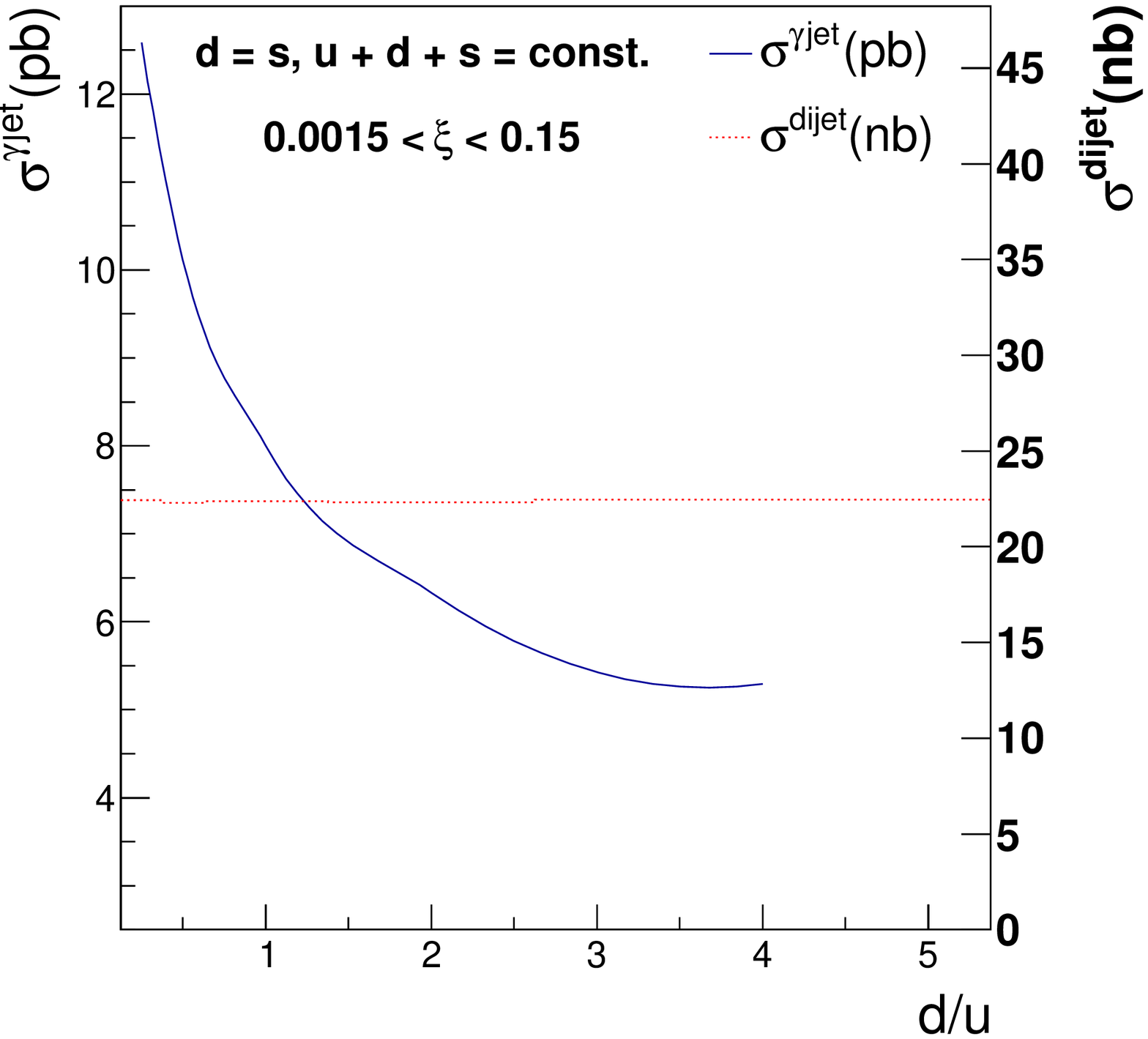,width=7cm}
\caption{DPE $\gamma +$ jet and di-jet cross section as a function of $d/u$, reflecting the Pomeron quark flavor composition. Left: acceptance of the 210m proton detectors. Right: acceptance of both the 210 and 420m detectors.}
\label{fig1}
\end{center}
\end{figure}

In this section, we detail the potential measurements to be performed at the LHC
in order to constrain for the first time the quark structure in the Pomeron.

\begin{figure}[t]
\begin{center}
\epsfig{file=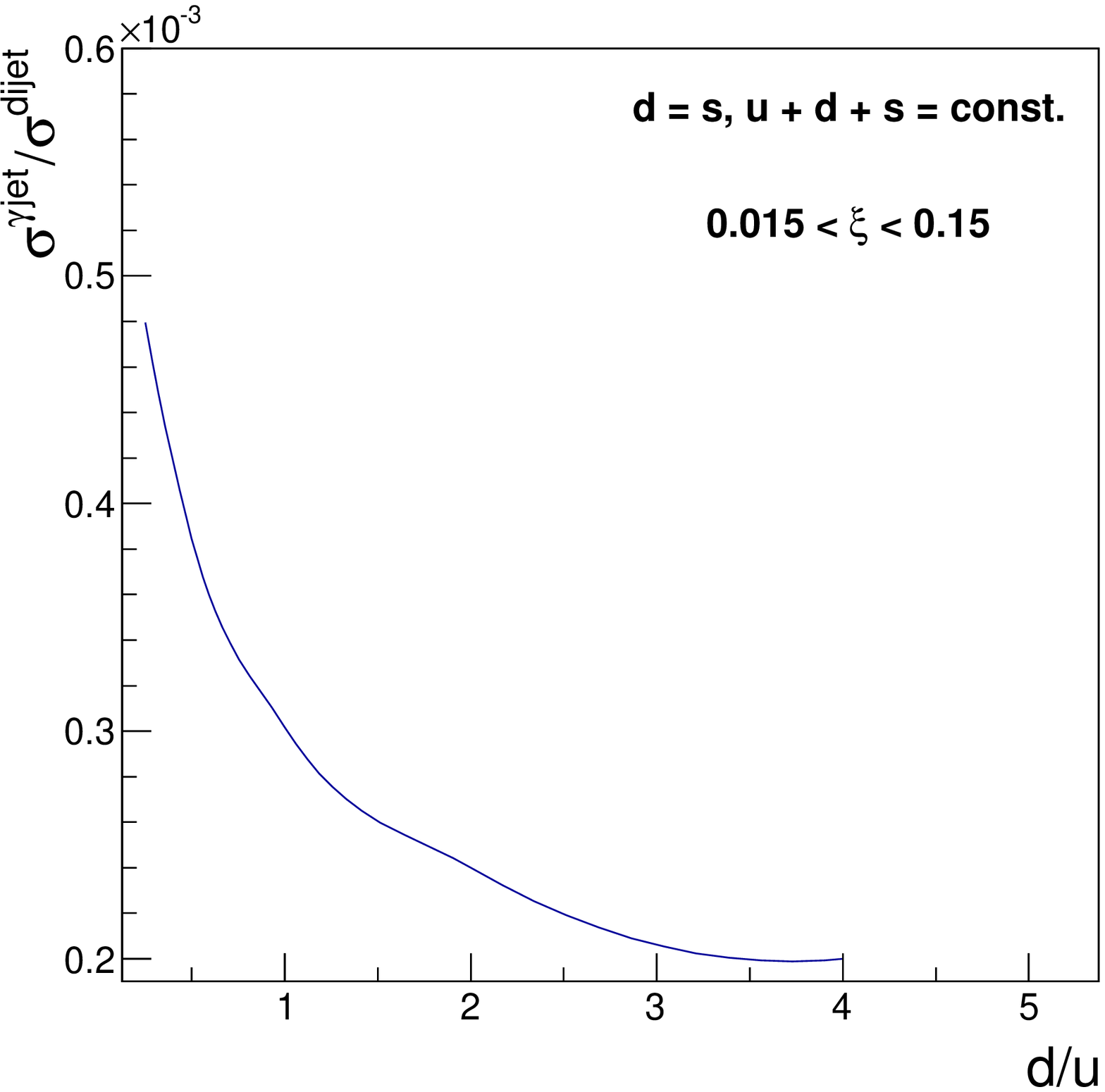,width=6.5cm}
\epsfig{file=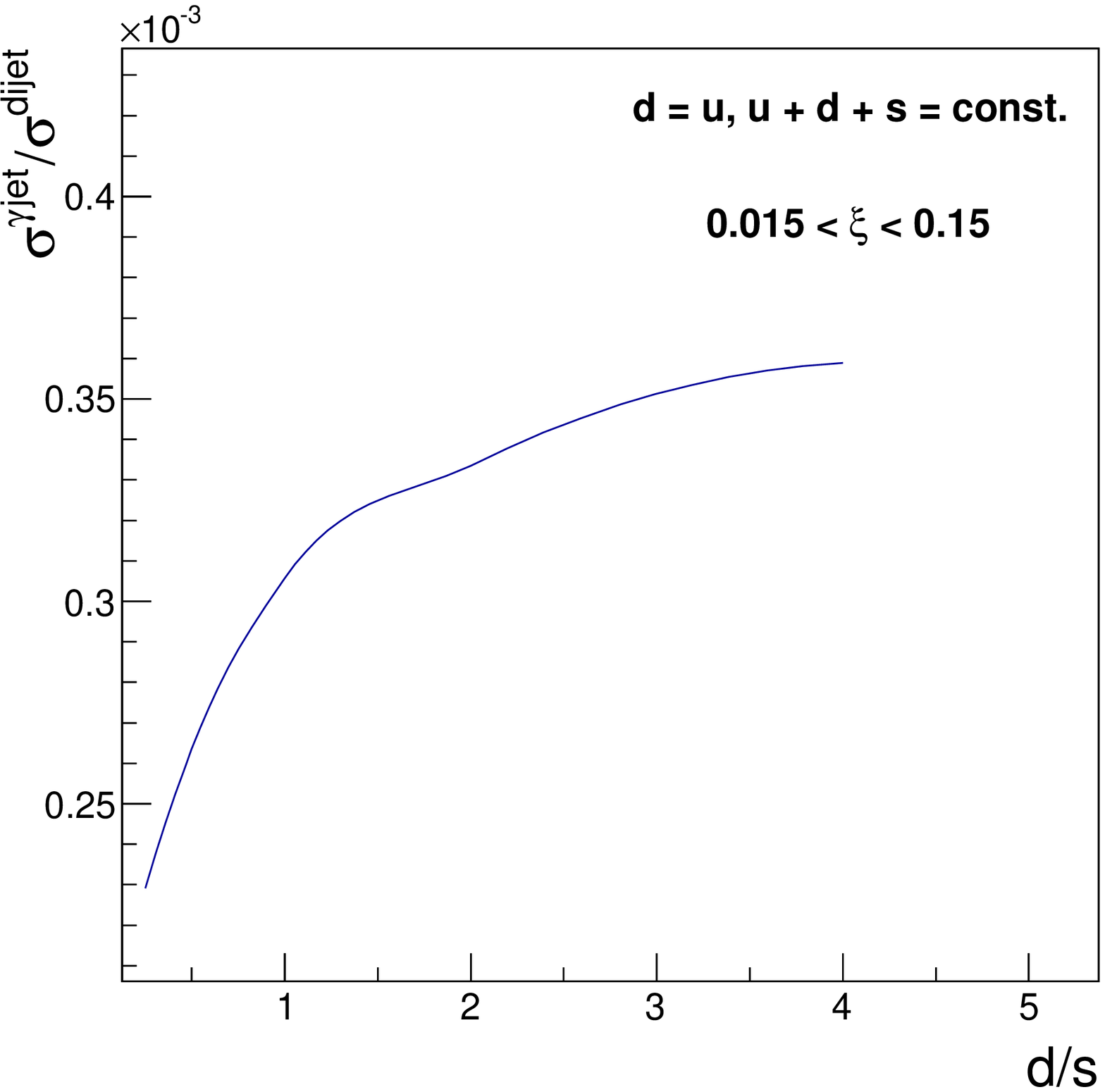,width=6.5cm}
\caption{DPE $\gamma +$ jet to di-jet cross section ratio, for the acceptance of the 210m proton detectors.
Left: as a function of $d/u$. Right: as a function of $d/s$.}
\label{fig2}
\end{center}
\end{figure}

Fig.~\ref{fig1} displays the $\gamma +$jet and dijet cross sections as a
function of the $d/u$ quark content of the Pomeron for two different acceptances
in $\xi$ for the forward detectors, $0.015<\xi <0,15$ (left) and $0.0015 < \xi
<0.15$ (right), while keeping $u+d+s$ constant. The $\gamma +$ jet cross section
varies by about a factor 2.5 when $u/d$ changes, while the dijet cross section 
remains constant
since it originates from gluon exchanges. As expected, the cross section for
$\gamma +$ jet events is much smaller and will limit the statistics of the
measurement. We note that a luminosity of 200-300 pb$^{-1}$ is needed to perform
the measurement with enough statistics when the proton is detected in AFP.
The main limiting factor is obviously the AFP acceptance since the diffractive
mass has to be larger than 350 GeV for 210 m detectors and about 100 GeV for 210
and 420 m detectors together. However, we will notice in the following that
detecting both protons is crucial for this measurement.
Fig.~\ref{fig2} displays the $\gamma +$ jet to the dijet cross section ratio as
a function of the $d/u$ quark content in the Pomeron for two different
acceptances in $\xi$ corresponding to the 210 and 210/420 m detectors. 
Measuring ratios is definitely better since most of systematics uncertainties 
will cancel. As expected, the ratios vary by factor 2.5 following the
assumptions on $u/d$ in the proton. 

\begin{figure}[!htb]
\begin{center}
\epsfig{file=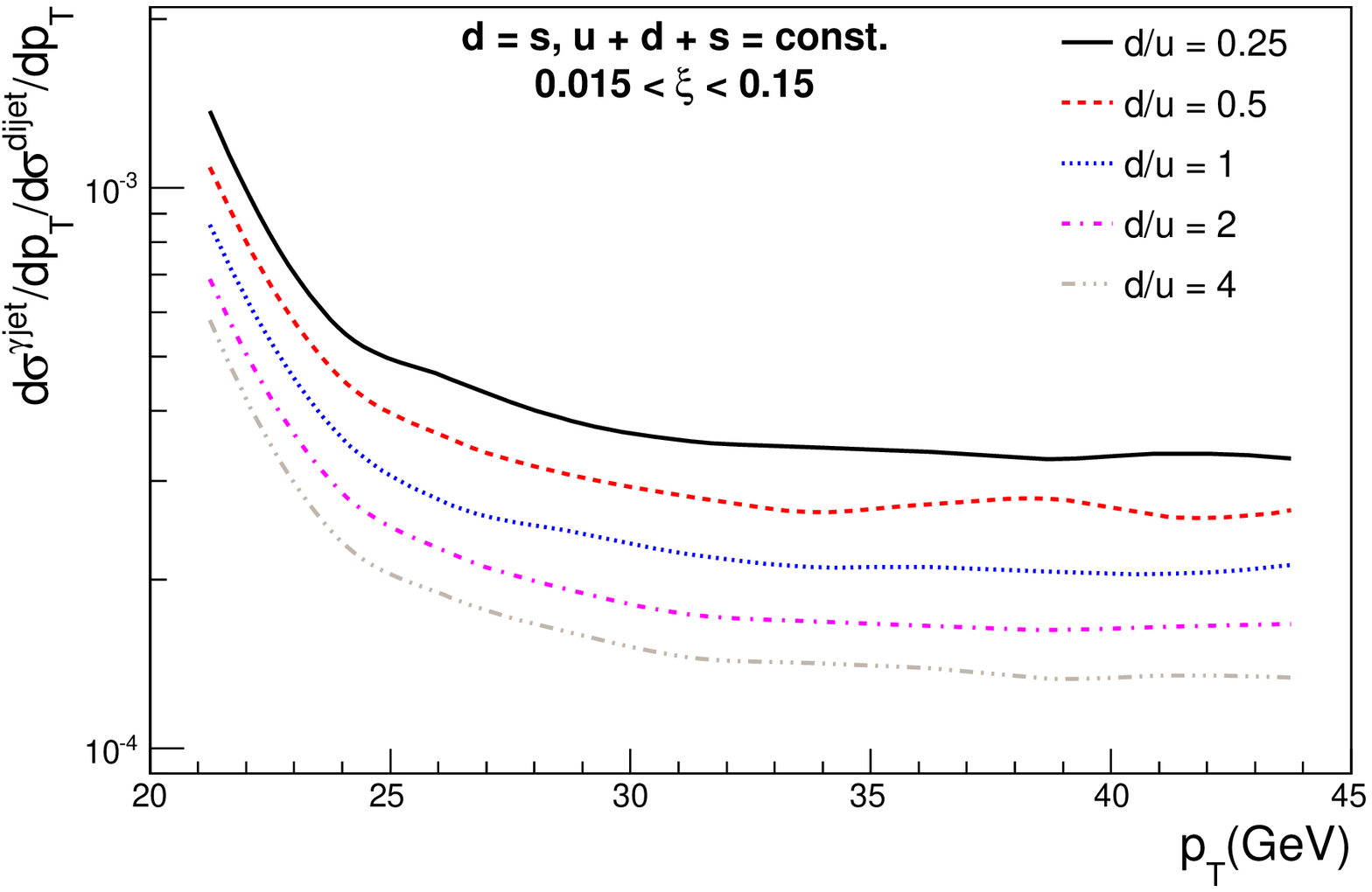,width=8.5cm}
\epsfig{file=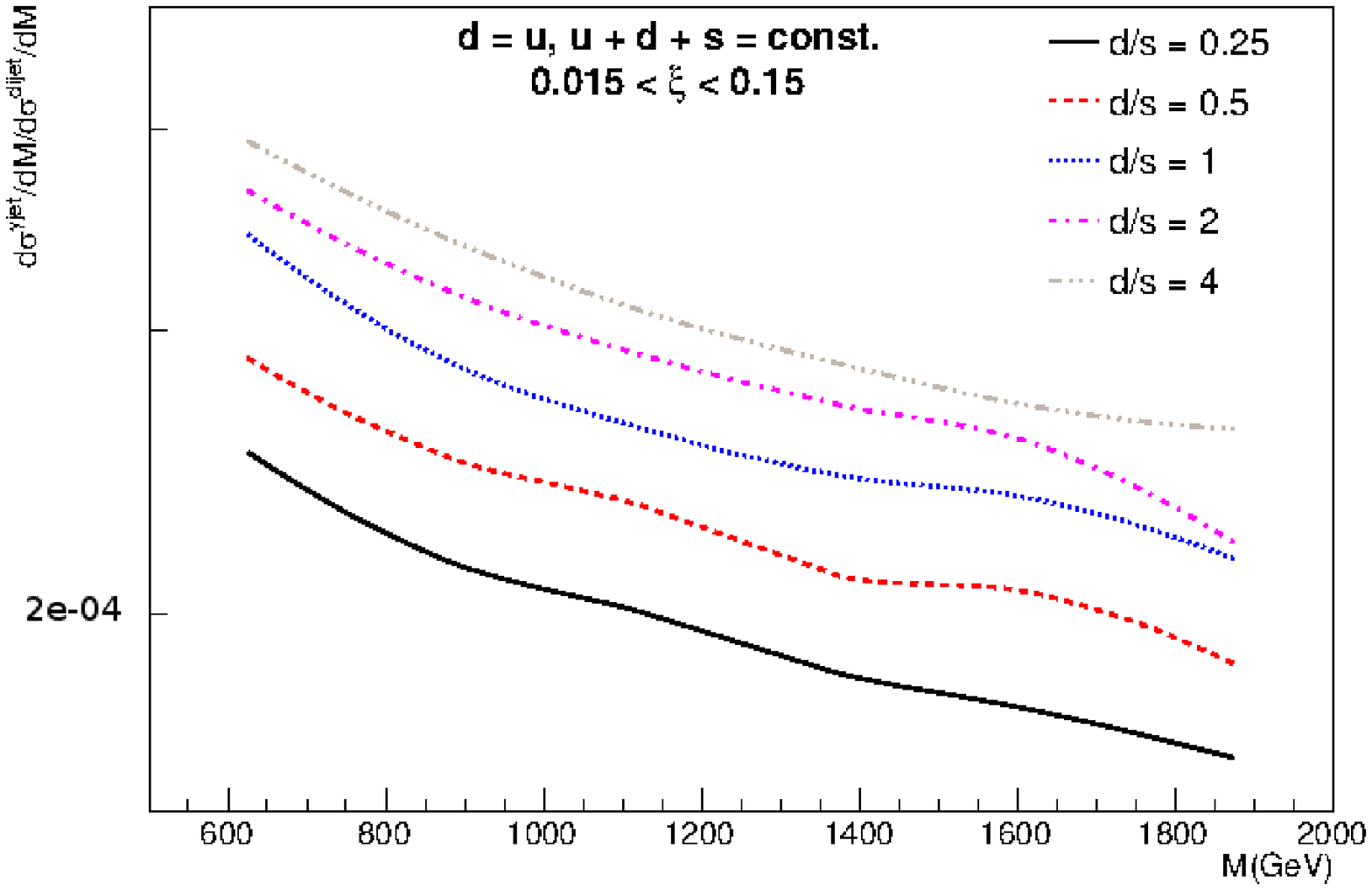,width=8.5cm}
\caption{DPE $\gamma +$ jet to di-jet differential cross section ratio, for the acceptance of the 210m proton detectors.
Left: as a function of  jet $p_T$, for different values of $d/u$. Right: as a function of the diffractive mass $M$, for different values of $d/s$.}
\label{fig3}
\end{center}
\end{figure}

\begin{figure}[!htb]
\begin{center}
\epsfig{file=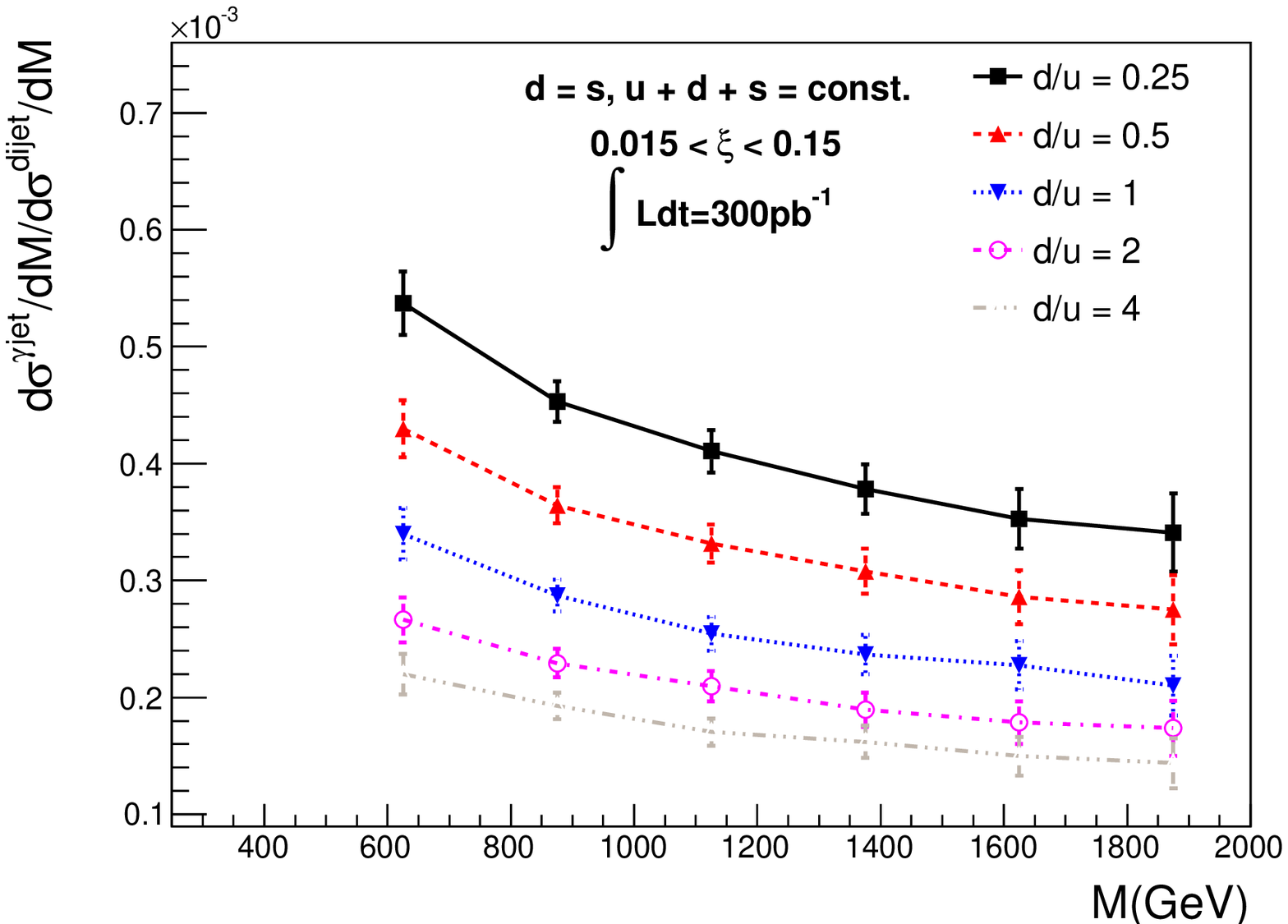,width=8.5cm}
\epsfig{file=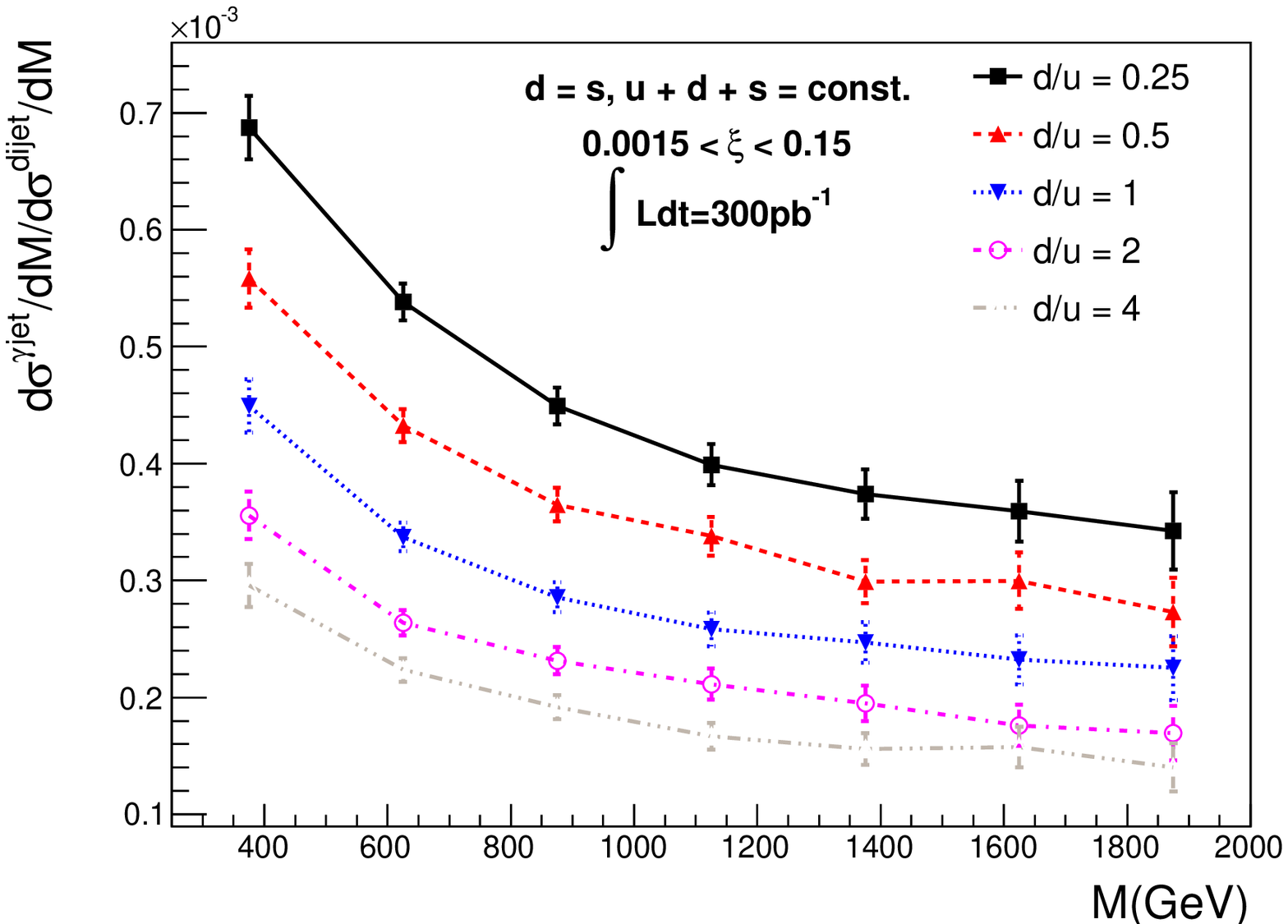,width=8.5cm}
\caption{DPE $\gamma +$ jet to di-jet differential cross section ratio as a function of the diffractive mass $M$, for different values of $d/u$.
Left: acceptance of the 210m proton detectors. Right: acceptance of both the 210 and 420m detectors.}
\label{fig4}
\end{center}
\end{figure}

Figs.~\ref{fig3} and \ref{fig4} display possible observables at the LHC that can 
probe the
quark content in the Pomeron. Fig.~\ref{fig3} displays the $\gamma +$jet to the
dijet cross section ratios as a function of the leading jet $p_T$ for different
assumptions on the quark content of the Pomeron, $d/u$ varying between 0.25 and
4 in steps of 0.25. We notice that the cross section ratio varies by a factor
2.5 for different values of $u/d$ and the ratio depends only weakly on the jet $p_T$
except at low values of jet $p_T$, which is due to the fact that we select
always the jet with the highest $p_T$ in the dijet cross section (and this is
obviously different for the $\gamma +$ jet sample where we have only one jet
most of the time).  
The aim of the jet $p_T$ distribution measurement is twofolds: is the Pomeron 
universal between
HERA and the LHC and what is the quark content of the Pomeron? As it was
mentionned in Section II, the QCD diffractive at HERA assumed that
$u=d=s=\bar{u}=\bar{d}=\bar{s}$, since data were not sensitive to the
difference between the different quark component in the Pomeron. 
The LHC data will allow us to determine for instance which value of $d/u$ is
favoured by data. Let us assume that $d/u=0.25$ is favoured. If this is the
case, it will be needed to go back to the HERA QCD diffractive fits and check if
the fit results at HERA can be modified to take into account this assumption. If
the fits to HERA data lead to a large $\chi^2$, it would indicate that the Pomeron is not
the same object at HERA and the LHC. On the other hand, if the HERA fits work
under this new assumption, the quark content in the Pomeron will be further
constrained. The advantage of measuring the cross section ratio as a function of
jet $p_T$ is that most of the systematic uncertainties due to the 
determination of the jet energy scale will cancel. This is however not the case
for the jet energy resolution since the jet $p_T$ distributions are different for
$\gamma +$jet and dijet events.

\begin{figure}[!htb]
\begin{center}
\epsfig{file=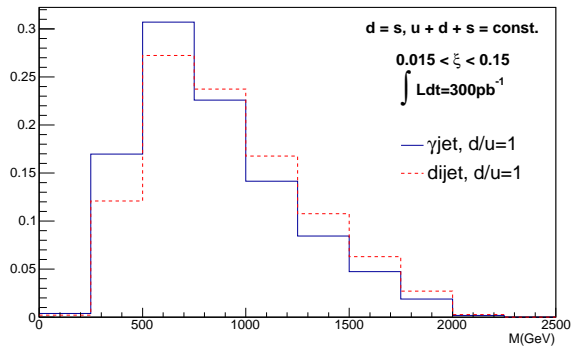,width=8.5cm}
\caption{Distribution (normalised to 1) of the total diffractive mass ($M=\sqrt{\xi_1 \xi_2 S}$)
for DPE production of dijets (dashed line, right scale) and $\gamma +$jet events 
(full line, left scale).
Since the distributions are similar the systematics uncertainties and the
resolution effects will almost cancel in the cross section ratio.}
\label{fig4b}
\end{center}
\end{figure}

Fig.~\ref{fig4} displays the $\gamma +$jet to dijet cross section ratio as a
function of the diffractive mass $M$ computed from the proton $\xi$ measured in the
forward detectors $M=\sqrt{\xi_1 \xi_2 S}$ where $\xi_1$ and $\xi_2$ are the momentum
fraction of the proton carried by each Pomeron and measured in the proton detectors. 
The advantage of this variable is
that most of systematic uncertainties due to the measurement of the diffractive
mass cancel since the mass distributions for $\gamma +$jet and dijet are similar
(see Fig.~\ref{fig4b}). The typical resolution on mass is in addition very good of the order
of 1 to 2\%. The statistical uncertainties corresponding to 300 pb$^{-1}$,
three weeks of data taking at low pile up, are also shown on the Figure. This
measurement will be fundamental to constrain in the most precise way the Pomeron
structure in terms of quark densities, and to test the Pomeron universality between
the Tevatron and the LHC.

\section{Comparison with soft color interaction models}

\subsection{Soft colour interaction models}
Soft color interaction models~(SCI)
describe~\cite{Edin:1995gi,Rathsman:1998tp}
additional interactions between colored partons
below the conventional cutoff for perturbative QCD.
These are based on the assumption of factorization between the conventional
perturbative event and the additional non-perturbative soft interactions.
The exchanges being soft means that the changes in momenta due to the additional exchanges
are very small, whereas the change in the event's color topology due to
exchanges of color charge can lead to significant observables,
e.g.\ rapidity gaps and leading beam remnants.
The probability to obtain a leading proton at the LHC in the context of
SCI models depends on the color charge
and the kinematic variables of the beam remnant before hadronization.
In string hadronization models, a colored remnant will be modeled as an
endpoint of a string, and the hadronization procedure will cause the
momentum of the beam remnant to be most likely distributed between mutliple
hadrons, so that a leading beam remnant on parton level will not be turned
into a single leading hadron.
On the other hand, a color-singlet remnant on parton level can be mapped
on a single leading hadron if the kinematics allow it.
To obtain a single leading proton within SCI models, the beam remnant may
therefore not been disturbed too much by the perturbative part of the event,
i.e.\ $\sqrt{t} \sim \Lambda_\textnormal{QCD}$.
This is seen as the typical experimentally observed distribution of
$e^{-bt}$ with $b \sim (2\Lambda_\textnormal{QCD})^{-1} \sim 7$ GeV$^{-2}$.

\begin{figure}
\begin{minipage}{0.47\linewidth}
\epsfig{file=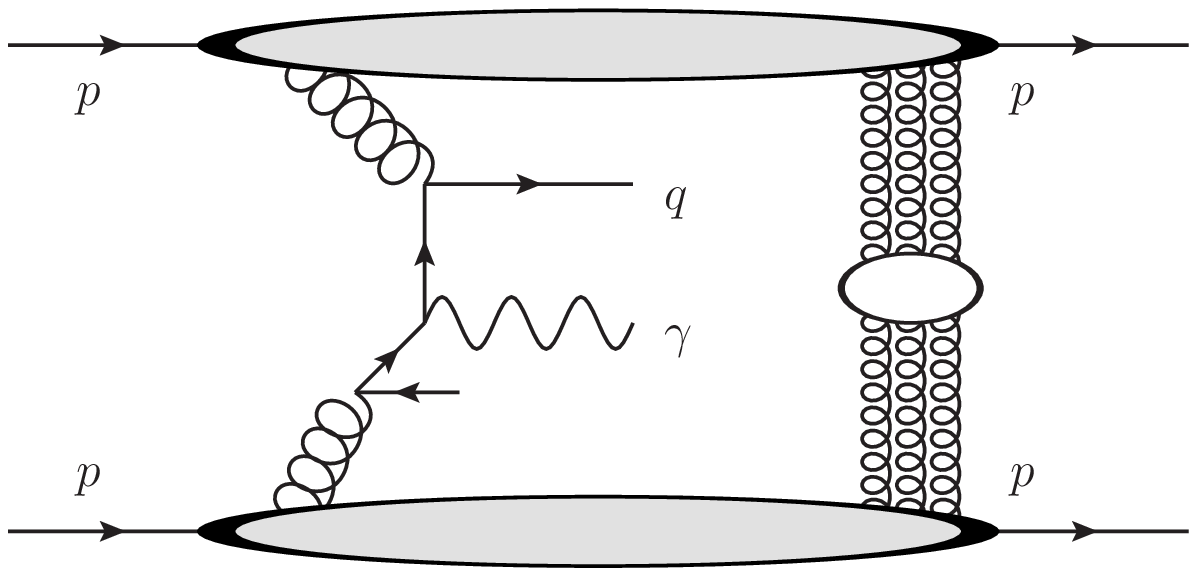,width=0.8\linewidth}
\caption{
$\gamma+$jet production described in terms of SCI.
This example depicts the case of two resolved gluons in the protons, one of which
branches into a $q\bar{q}$ pair, described by initial state radiation in a monte carlo program.
Matrix element for $qg \to q\gamma$ in the center.
Additional soft color exchanges depicted at the right end.
}
\label{fig:sci-feynman-gammajet}
\end{minipage}
\hspace{0.04\linewidth}
\begin{minipage}{0.47\linewidth}
\epsfig{file=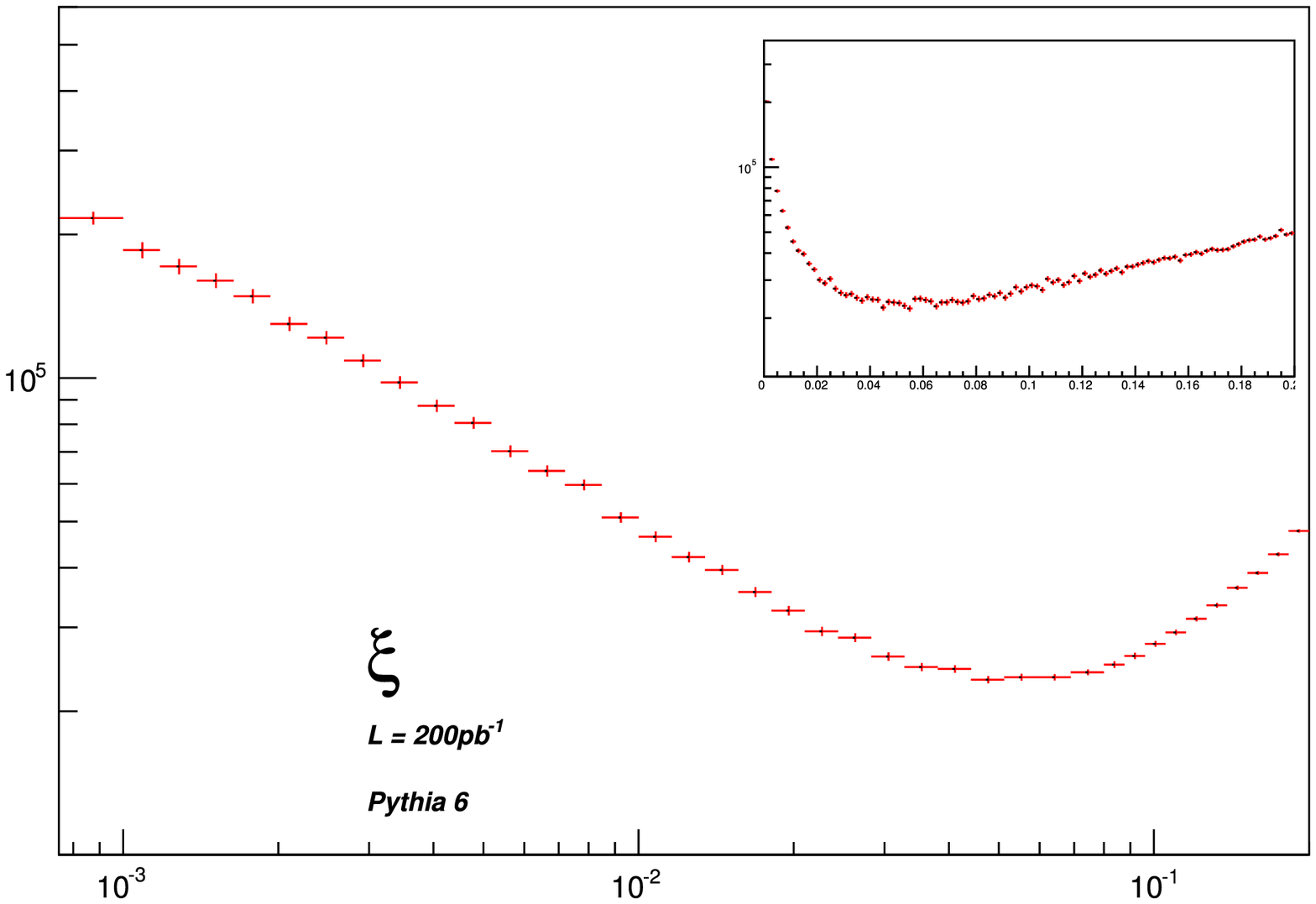,width=\linewidth}
\caption{
Distribution in $\xi$ of forward protons at $pp$ 14 TeV
for the SCI model requiring dijets with $p_T>20$~GeV.
}
\label{fig:xi_simple}
\end{minipage}
\end{figure}

A main conceptional difference between the DPE and SCI models is that the SCI
model treats ``diffractive'' and ``non-diffractive'' events on equal footing.
This implies in particular that a event sample
based on a cut in $\xi$ of the proton's momentum loss can also include events
where the beam remnant was not a color-singlet on parton level but
where the cut in $\xi$ was fulfilled by a leading proton from string hadronization.
Fig.\ref{fig:xi_simple} shows the distribution of events in the fractional
momentum loss $\xi$ of the proton on a single side for dijet events with
$p_T>20$~GeV.
The distribution shows the increase in cross section for very small momentum
losses of the beam protons, typical for diffractive scattering.
In order to compare~\cite{Ingelman:2012wj} with models which describe diffractive scattering
in terms of a Pomeron without additional Reggeons, it is therefore important
to constrain the analysis to the region of phase space where the momentum loss
of the leading final state protons is very small, so that the background due
``non-diffractive'' events can be neglected.
Treating diffractive and non-diffractive events on the same footing is a drawback
for the computational performance of the SCI implementation, since the
monte carlo integration is computationally very intensive.

\begin{figure}
\begin{minipage}{0.47\linewidth}
\epsfig{file=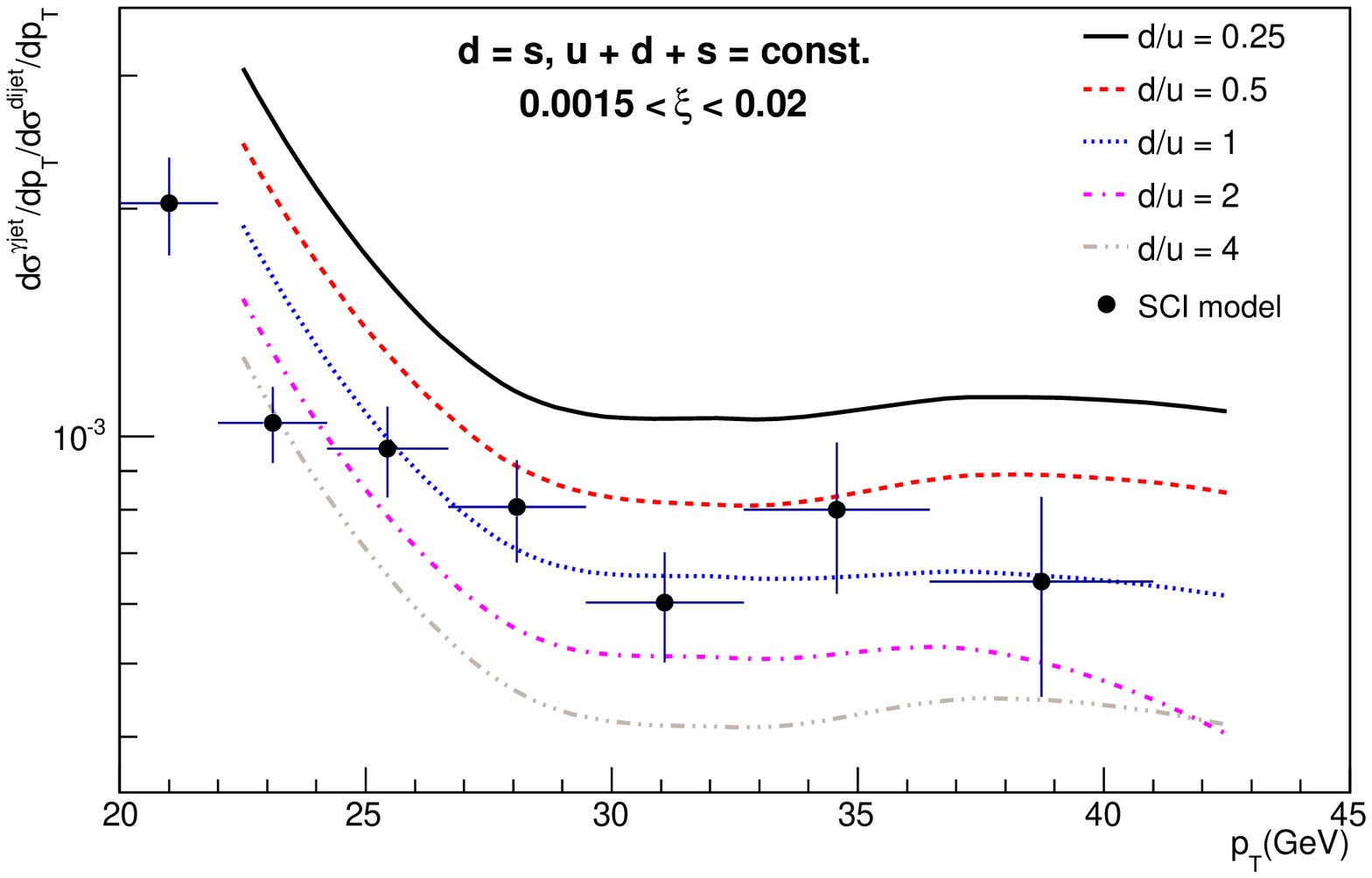,width=\linewidth}
\caption{
Ratio of $\gamma+$jet over jet$+$jet cross section
differential in the highest $p_T$ jet.
}
\label{fig:scidpe-m}
\end{minipage}
\hspace{0.04\linewidth}
\begin{minipage}{0.47\linewidth}
\epsfig{file=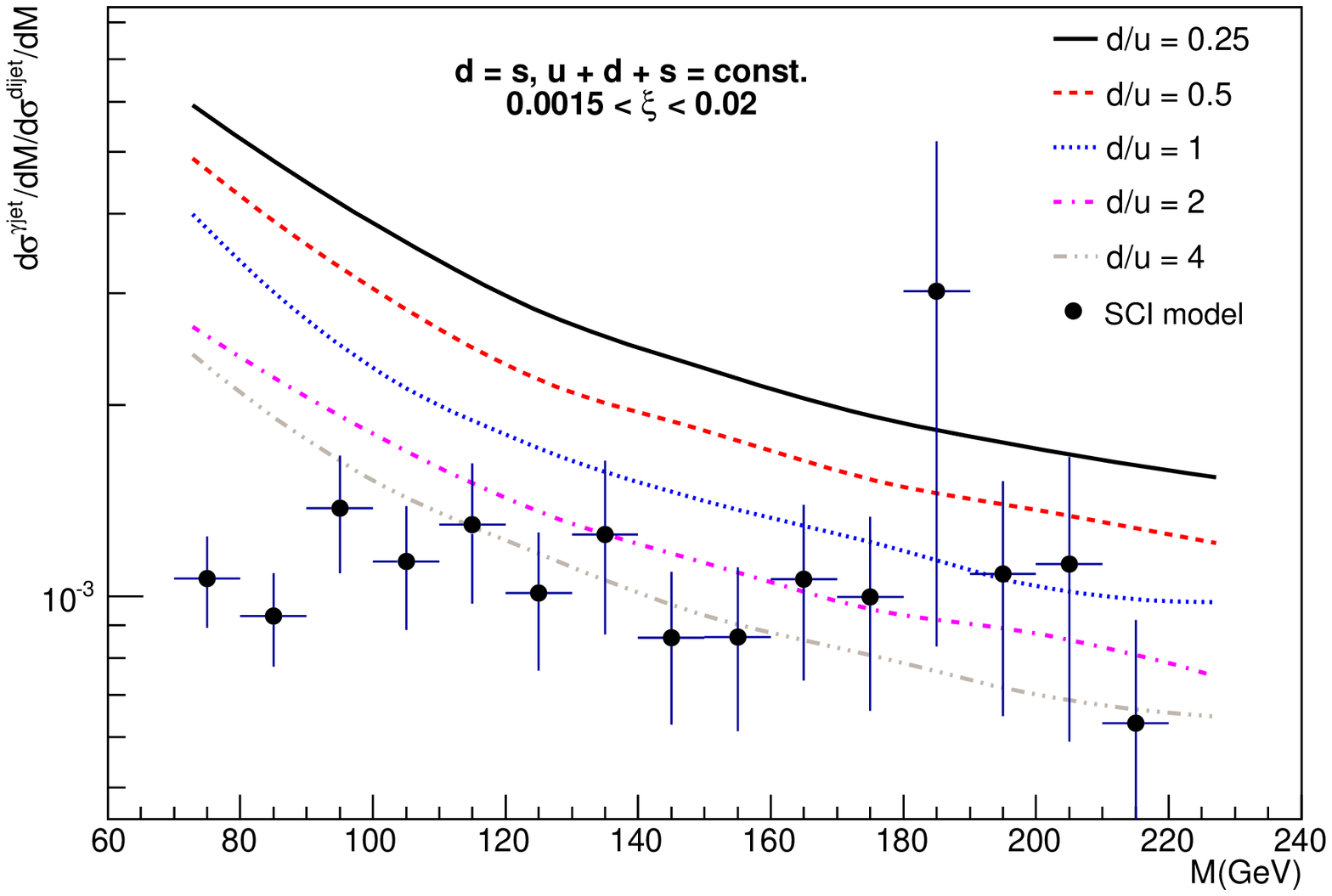,width=\linewidth}
\caption{
Ratio of $\gamma+$jet over jet$+$jet cross section
differential in the total diffractive mass $\sqrt{\hat{s}}=\sqrt{s \xi_1 \xi_2}$
given by the leading protons.
}
\label{fig:scidpe-sqrtshat}
\end{minipage}
\end{figure}

\subsection{$\gamma+$jet over jet$+$jet ratio from SCI}

In order to compare SCI and Pomeron like models, we request in the following
$\xi \le 0.02$ for the protons on both sides
in order to stay in the region of the diffractive peak in the SCI model,
even though a comparison with the $\xi$ distribution from CDF~\cite{cdf}
shows that the transition to a ``diffractive'' sample may be defined already at larger
values of $\xi$ up to about 0.1.
Predictions for the $\gamma$+jet and jet$+$jet cross sections are obtained
from Pythia 6.4 plus SCI with jet reconstruction via an anti-$k_T$ algorithm~\cite{anti-kt}.
An example for a basic process contributing to the $\gamma+$jet cross section is depicted
in Fig.\ref{fig:sci-feynman-gammajet},
where on both sides a gluon is resolved in the proton.
While one gluon can be directly connected to the hard matrix element,
the other gluon fluctuates in this example into a $q\bar{q}$ pair, giving the quark
for the hard process.
This illustrates another difference between the models,
which is the QCD evolution of the initial state,
treated like in standard non-diffractive processes,
whereas the Herwig/DPE model describes the process by just a hard matrix element
which resolves the content of the Pomeron.
As before, we require in addition to the leading protons
$p_{T}>20$~GeV for the jets and the $\gamma$,
and $|\eta| < 4.4$.
These requirements translate for a central event at LHC 14 TeV to a minimum
$\xi \simeq 2.9 \times 10^{-3}$.
The contributing hard matrix elements for events passing the cuts
are for the $\gamma+$jet cross section
$fg \to f\gamma$ (90\%) and
$f\bar{f} \to g\gamma$ (5\%),
and for the jet$+$jet cross section
$gg \to gg$ (50\%),
$fg \to fg$ (38\%),
$gg \to f\bar{f}$ (8\%) and
$f\bar{f} \to f\bar{f}$ ($\sim$2\%).
We compare in Figs.\ref{fig:scidpe-m} and \ref{fig:scidpe-sqrtshat}
the ratios of $\gamma+$jet and jet$+$jet
as distributions in mass of the central system $m$,
and leading jet $p_T$.
We note first the overall quite good agreement between both models,
which is remarkable especially because the SCI model's parameters
do not change between HERA, Tevatron and LHC,
and because Pythia/SCI uses the standard proton PDF without further modifications.
Fig.\ref{fig:scidpe-m} suggests best agreement of the Herwig/DPE model with Pythia/SCI
for $d/u \simeq 1$.
Fig.\ref{fig:scidpe-sqrtshat} shows that the models predict different
ratios of the cross sections when approaching smaller jet $p_T$.
In the case of Herwig/DPE, the ratio increases significantly towards smaller values
of $p_T$, whereas it is less dependent on diffractive mass for the case of
Pythia/SCI within the range of $\xi<0.02$ under consideration.
Allowing a more loose cut $\xi<0.03$, the ratio $R(\sqrt{\hat{s}})$ at a 
somewhat larger
value of $\sqrt{\hat{s}}$ behaves as
$R(100\textnormal{\:GeV})/R(400\textnormal{\:GeV}) \simeq 1.8$
showing a slow decrease with $p_T$. By measuring
the distribution of the $\gamma +$jet to the dijet cross section ratio,
it will be possible to
distinguish for the first time between SCI/Pythia and DPE/Herwig models 
using in particular the slope of the
diffractive mass distribution. Let us note further that the MC predictions
depend also on the specific tune of the Monte Carlo and we indeed test the
prediction of the model as implemented in a Monte Carlo. In addition,
it is worth noticing that the comparison between both
models can surely be accomplished up to $\xi \sim$0.1 (if we follow the CDF
data) and thus in the acceptance of the 210 m forward proton detectors.

\section{Conclusion}
The measurement of diffractive processes at the LHC will allow us to get a
better understanding of the Pomeron. The measurement of the dijet cross section
with a luminosity of 10 pb$^{-1}$ (one day of data taking at low luminosity at
the LHC) increases our knowledge on the gluon content in the Pomeron and allows
us to check if the Poemron is indeed the same at $ep$ or $pp$ colliders, which
is a fundamental theoretical question. This measurement will be made possible by
the installation of forward proton detectors located at about 210 m (and 420 m)
from the interaction point of the ATLAS and CMS collaborations. 

The measurement of the ratio of the $\gamma +$jet to the dijet cross section
allows us to assess directly the quark content of the Pomeron (let us recall
that present QCD fits assume $u=\bar{u}=d=\bar{d}=s=\bar{s}$) and again to
compare the structure of the Pomeron at $ep$ and $pp$ colliders in addition of
constraining the quark structure in the Pomeron. The best observable is the
total diffractive mass obtained from the reconstruction of the proton momentum
loss measured in the forward proton detectors, since most of the systematics
disappear in the ratio of the $\gamma+$jet to the dijet cross sections.

In addition, diffractive processes with two identified leading outgoing protons
allows us to test the predictions of soft color exchange models at unprecedented
center-of-mass energies.
We find an overall good agreement between Herwig/DPE and Pythia/SCI
for the prediction of the ratio between $\gamma+$jet and jet$+$jet cross sections,
but the distribution of this ratio as a function of the total
diffractive mass distributions
may allow to distinguish between the DPE/Herwig and SCIO/Pythis models, giving further 
insight into soft QCD. For the first time, it seems a good method to probe the
differences between SCI/Pythia and Pomeron/Heriwg models
at hadronic colliders since SCI/Pythis
models lead to a flatter dependence of the total diffractive mass.

\begin{acknowledgments}

We thank G. Ingelman, O. Kepka, M. Mangano and R. Peschanski for discussions and comments.

\end{acknowledgments}

\end{document}